\documentclass[12pt]{article}
\usepackage{amscd,amssymb,amsmath,latexsym,enumerate}
\usepackage[mathscr]{euscript}
\usepackage{epsfig}
\usepackage{fancybox}
\usepackage{verbatim}

\usepackage{color}

\title{Parity as $\ZM_2$-valued spectral flow}

\author{Nora Doll$^1$, Hermann Schulz-Baldes$^1$ and Nils Waterstraat$^2$
\\
\\
{\small $^1$ Department Mathematik,  Friedrich-Alexander-Universit\"at Erlangen-N\"urnberg, }
\\
{\small Cauerstr. 11, 91058 Erlangen, Germany}
\\
{\small $^2$ Institut f\"ur Mathematik, Martin-Luther-Universit\"at Halle-Wittenberg, }
\\
{\small Theodor-Lieser-Str. 5, 06120 Halle, Germany}
}

\date{ }

\newtheorem{theo}{Theorem}
\newtheorem{defini}{Definition}
\newtheorem{proposi}{Proposition}
\newtheorem{lemma}{Lemma}
\newtheorem{coro}{Corollary}

\newcommand{\CM}{{\mathbb C}}
\newcommand{\NM}{{\mathbb N}}
\newcommand{\RM}{{\mathbb R}}

\newcommand{\ZM}{{\mathbb Z}}
\newcommand{\PM}{{\mathbb P}}

\newcommand{\Ee}{{\cal E}}

\newcommand{\Bb}{{\cal B}}

\newcommand{\Ff}{{\cal F}}

\newcommand{\Oo}{{\cal O}}

\newcommand{\Cc}{{\cal C}}

\newcommand{\Qq}{{\cal Q}}
\newcommand{\Kk}{{\cal K}}
\newcommand{\Hh}{{\cal H}}

\newcommand{\one}{{\bf 1}}
\newcommand{\proj}{p}

\newcommand{\SF}{{\rm Sf}} 
\newcommand{\PI}{{\rm Ind}_2} 
\newcommand{\PF}{{\sigma}} 
 
\newcommand{\Ind}{{\rm Ind}} 
\newcommand{\Ker}{{\rm Ker}} 
\newcommand{\degLS}{{\rm deg}_2} 
\newcommand{\Ran}{{\rm Ran}} 
\newcommand{\sgn}{{\rm sgn}} 
 
\newcommand{\diag}{{\rm diag}} 
\newcommand{\spec}{{\rm spec}}

\newcommand{\isom}{V}

\begin{document}

\maketitle

\begin{abstract}
This note is about the topology of the path space of linear Fredholm operators on a real Hilbert space. Fitzpatrick and Pejsachowicz introduced the parity of such a path, based on the Leray-Schauder degree of a path of parametrices. Here an alternative analytic approach is presented which reduces the parity to the $\ZM_2$-valued spectral flow of an associated path of chiral skew-adjoints. Furthermore the related notion of $\ZM_2$-index of a Fredholm pair of chiral complex structures is introduced and connected to the parity of a suitable path. Several non-trivial examples are provided. One of them concerns topological insulators, another an application to the bifurcation of a non-linear partial differential equation.
\hfill
\end{abstract}


\section{Introduction}

The spectral flow for paths of self-adjoint Fredholm operators on a complex Hilbert space is a well-known homotopy invariant \cite{APS,Ph,FP0,Les}. It plays a role in numerous other fields, {\it e.g.} index theory \cite{APS,Ph1,CP,DS2} and bifurcation theory \cite{FP0,FP}. For $\RM$-linear operators on a real Hilbert space $\Hh_\RM$, spectral flow is still a well-defined and useful object. Moreover, for paths $[0,1]\ni t\mapsto B_t$ of arbitrary (not necessarily self-adjoint) Fredholm operators on $\Hh_\RM$ a $\ZM_2$-valued parity $\sigma$ has been introduced by Fitzpatrick and Pejsachowicz \cite{FP0},  and for paths $[0,1]\ni t\mapsto T_t$ of skew-adjoint real Fredholm operators a $\ZM_2$-valued spectral flow $\SF_2$ has also been studied \cite{CPS}. This note presents the parity of a path $[0,1]\ni t\mapsto B_t$ of real Fredholm operators as the $\ZM_2$-valued spectral flow of an associated path of chiral skew-adjoint Fredholm operators on $\Hh_\RM \oplus \Hh_\RM$:
\begin{equation}
\label{eq-SigmaSF2}
\sigma\big([0,1]\ni t\mapsto B_t\big)
\;=\;
\SF_2
\left([0,1]\ni t\mapsto
\begin{pmatrix} 0 & B_t \\ -B_t^* & 0
\end{pmatrix}
\right)
\;.
\end{equation}
This provides a new perspective on parity and also allows to deduce its main properties directly from known facts on the $\ZM_2$-valued spectral flow. We also believe that the presented approach makes the parity more accessible for computations.  A new result for the parity is an index formula for paths between conjugate Fredholm pairs of complex structures, see Section~\ref{sec-index}. This corresponds to analogous results for the spectral flow between conjugate Fredholm pairs of projections \cite{Ph1} as well as the $\ZM_2$-valued spectral flow \cite{CPS}. 

\vspace{.2cm}

To further stress the similarities between spectral flow, $\ZM_2$-valued spectral flow and parity, let us consider the classifying spaces for real $K$-theory as introduced by Atiyah and Singer \cite{AS}. Let $\Ff^k=\Ff^k(\Hh_\RM)$ denote the space of skew-adjoint Fredholm operators on a real separable Hilbert space $\Hh_\RM$ which anticommute with representations  $I_1,\ldots,I_{k-1}$ of the generators of a real Clifford algebra of signature $(0,k-1)$ \cite{AS}.  By reducing out these relations in a concrete representation, it is possible (but tedious) to identify each $\Ff^k$ with a set of Fredholm operators on $\Hh_\RM$ having certain supplementary symmetry relations. Relevant for the following is that $\Ff^0\cong\Ff^8$ is isomorphic to the set of all Fredholm operators on $\Hh_\RM$, $\Ff^1$ is isomorphic to the set of skew-adjoint Fredholm operators while $\Ff^7$ is isomorphic to the self-adjoint Fredholm operators on $\Hh_\RM$ with positive and negative essential spectrum. Furthermore, $\Ff^3$ is isomorphic to the set of those elements of $\Ff^7$ that are linear over the quaternions. Atiyah and Singer \cite{AS} found that the homotopy groups of these spaces satisfy
$$
\pi_{j}(\Ff^i)
\;=\;
\pi_{0}(\Ff^{i+j})
\;=\;
\pi_{j+i}(\Ff^0)
\;,
$$
and are given explicitly by 
\begin{equation}
\label{eq-homotopygroups}
\begin{tabular}{|c||c|c|c|c|c|c|c|c|}
\hline
$i$ & $0$ & $1$ & $2$ & $3$ & $4$ & $5$ & $6$ & $7$  \\
\hline
$\pi_{0}(\Ff^i)$  &$\ZM$ & $\ZM_2$ & $\ZM_2$ & 0 & $2\,\ZM$ & $0$ & $0$ & $0$ 
\\
\hline
$\pi_{1}(\Ff^i)$  &$\ZM_2$ & $\ZM_2$ & $0$ & $2\,\ZM$ & $0$ & $0$ & $0$ & $\ZM$ 
\\
\hline
\end{tabular}
\end{equation}
The components $\pi_0(\Ff^i)$ in the second row are labelled by the index (for $i=0,4$) and the $\ZM_2$-index is given by the nullity modulo $2$ (for $i=1,2$). The spectral flow provides an explicit isomorphism from $\pi_{1}(\Ff^7)$ to $\ZM$, and also from $\pi_{1}(\Ff^3)$ to $2\,\ZM$. Here the factor $2$ merely stresses that eigenvalues of self-adjoint quaternionic operators are always of even multiplicity. More precisely, if a self-adjoint quaternionic matrix is written as a complex matrix of double size, then this complex matrix has a symmetry leading to even dimensional eigenspaces (just like time reversal for fermions with half-integer spin leads to Kramers' degeneracy). Therefore also the spectral flow along paths of quaternionic operators is even. Furthermore the parity gives the isomorphism $\pi_1(\Ff^0)\cong\ZM_2$ \cite{FP} and the $\ZM_2$-valued spectral flow provides the isomorphism $\pi_1(\Ff^1)\cong\ZM_2$ \cite{CPS}. Hence the spectral flow, parity and $\ZM_2$-valued spectral flow allow to detect the topology in the last row of \eqref{eq-homotopygroups}. Furthermore, in view of table \eqref{eq-homotopygroups}, one does not expect there to be any other flow of interest. Let us also note that \eqref{eq-SigmaSF2} results from realizing $\Ff^0$ as those elements of $\Ff^1$ that anticommute with the representation $J$ of the generator of a real Clifford algebra of signature $(1,0)$. Explicitly, $J=\diag(\one,-\one)$ in its spectral representation and elements $T\in \Ff^1$ with the so-called chiral symmetry $JTJ=-T$ are off-diagonal as on the right hand-side of \eqref{eq-SigmaSF2}. This reduction is in the opposite direction than the one considered in \cite{AS}. Moreover, chiral skew-adjoints often also appear in different guise in applications. An  example are chiral self-adjoints, see Section~\ref{sec-Extensions}, which are naturally associated to chiral topological insulators, see Section~\ref{sec-TopIns}. Finally, let us stress that while table \eqref{eq-homotopygroups} only concerns closed loops, the definition of spectral flow, $\ZM_2$-valued spectral flow and parity apply to arbitrary (open) paths. 

\vspace{.2cm}

In Section~\ref{sec-PI} a $\ZM_2$-index of a Fredholm pair of chiral complex structures is introduced. This is the parity version of Kato's index of a Fredholm pair of projections \cite{Kat} as further studied by Avron, Seiler and Simon \cite{ASS}. This is closely tied to the parity, as explained in Section~\ref{sec-PI} and of particular interest and importance for Fredholm pairs given by unitary conjugates. This leads to an index formula proved in Section~\ref{sec-index}. Finally Sections~\ref{sec-TopIns} and \ref{sec-Bifur} give two applications of the parity.

\section{Parity in finite dimension}

The characterizing features of the parity can best be understood in finite dimension. Hence let us consider a (continuous) path $[0,1] \ni t \mapsto B_t$ of real $N\times N$ matrices acting on the real Hilbert space $\Hh_\RM=\RM^N$. Furthermore, let the path be admissible in the sense that its endpoints $B_0$ and $B_1$ are invertible, namely are in the general linear group $\mbox{\rm Gl}(N,\RM)$. This group has two components, specified by either a positive or a negative determinant. The parity  of the path $[0,1]\ni t \mapsto B_t$ is simply $1$ if the endpoints are in the same component and $-1$ if they are in the two different components. The following provides an analytic formula for this.

\begin{defini}
\label{def-PFfinite}
For an admissible path $[0,1]\ni t \mapsto B_t$ of real $N\times N$ matrices, the parity is defined as
\begin{equation}
\label{eq-PFfinite}
\PF([0,1] \ni t \mapsto B_t)
\;=\;
\sgn(\det(B_1))\,\sgn(\det(B_0))\;\in\;\ZM_2
\;,
\end{equation}
where $\ZM_2$ is viewed as the multiplicative group $\ZM_2=\{-1,1\}$. As this only depends on the endpoints, we will also simply write $\PF(B_0,B_1)$. After rescaling, all of this also applies to paths $[a,b]\ni t \mapsto B_t$ with arbitrary endpoints $a<b$.
\end{defini}

The definition directly implies that the parity $\PF$ of admissible paths of real matrices is a homotopy invariant (under homotopies of the path keeping the endpoints fixed), it has a concatenation property and it is normalized in the sense that the parity of a path in the invertibles is $1$. Furthermore, one has a multiplicativity property under direct sums, namely for another admissible path $[0,1]\ni t \mapsto B_t'$ of real $L\times L$ matrices, the definition directly implies that
$$
\PF([0,1] \ni t \mapsto B_t\oplus B_t')
\;=\;
\PF([0,1] \ni t \mapsto B_t)
\;\cdot\;
\PF([0,1] \ni t \mapsto B_t')
\;,
$$
with multiplication in $\ZM_2$. 

\vspace{.2cm}

For the generalization to infinite dimension there are several possibilities \cite{Les}. The route taken by Fitzpatrick and Pejsachowicz~\cite{FP} uses the fact that $\sgn(\det(B))$ can, under suitable conditions, be extended to infinite dimensions as the Leray-Schauder degree, for details see Section~\ref{sec-InfDim} below. In this note we elaborate on another possibility which consists in first rewriting Definition~\ref{def-PFfinite} in terms of skew-adjoint matrices on a doubled Hilbert space, just as suggested by Atiyah and Singer \cite{AS}. This has the advantage that tools from the spectral analysis of skew-adjoint operators can be used and the connection to the $\ZM_2$-valued spectral flow from \cite{CPS}  is uncovered. Hence let us use the real Hilbert space $\Hh'_\RM=\Hh_\RM\oplus\Hh_\RM$ equipped with the $\ZM_2$-grading $J=\diag(\one,-\one)$. Set:
\begin{equation}
\label{eq-link}
T_t\;=\;\begin{pmatrix} 0 & B_t \\ -B_t^* & 0 \end{pmatrix}
\;.
\end{equation}
These operators have a so-called chiral symmetry:
\begin{equation}
\label{eq-ChiralSym}
J\,T_t\,J\;=\;-T_t
\;.
\end{equation}
Conversely, if one has a real Hilbert space $\Hh'_\RM$ equipped with the $\ZM_2$-grading given by a self-adjoint unitary $J=J^*=J^{-1}$ and a path $[0,1]\ni t\mapsto T_t$ of real chiral skew-adjoints, then going to the spectral representation of $J$ in which $J=\diag(\one,-\one)$ leads to the representation of $T_t$ in the form \eqref{eq-link}. Hence \eqref{eq-link} provides a bijection between the set of paths of operators on $\Hh_\RM$ and the set of paths of chiral skew-adjoints on $\Hh'_\RM$. The chiral symmetry \eqref{eq-ChiralSym} implies that the spectrum always satisfies $\spec(T_t)=-\spec(T_t)\subset \imath\,\RM$. A non-trivial topology in the path is detected by the $\ZM_2$-valued spectral flow \cite{CPS}, the definition of which we recall next. For this purpose, let us note that the endpoints $T_0$ and $T_1$ are invertible (because the initial path was admissible) and therefore there exists an invertible $A$ such that $T_1=A^*T_0A$. Then, by definition \cite{CPS}, 
\begin{equation}
\label{eq-PFfinite2}
\SF_2([0,1]\ni t \mapsto T_t)
\;=\;
\sgn(\det(A))\;\in\;\ZM_2
\;.
\end{equation}
As the definition of $\SF_2([0,1]\ni t \mapsto T_t)$ only depends on the endpoints we will also write $\SF_2(T_0, T_1)$. For $T_0$ and $T_1$ in the form \eqref{eq-link} one has $T_1=A^*T_0A$ for $A=\diag((B_0^*)^{-1}B_1^*,\one)$. This directly implies
\begin{equation}
\label{eq-PFfinite3}
\PF([0,1] \ni t \mapsto B_t)
\;=\;
\SF_2([0,1]\ni t \mapsto T_t)
\end{equation}
whenever the identification \eqref{eq-link} holds. This explains why \eqref{eq-SigmaSF2} holds in finite dimension.  The $\ZM_2$-valued spectral flow given by \eqref{eq-PFfinite2} has an invariance property under conjugation, namely if $[0,1]\ni t \mapsto O_t$ is a path of orthogonals commuting with $J$, then
$$
\SF_2([0,1]\ni t \mapsto O_tT_tO_t^*)
\;=\;
\SF_2([0,1]\ni t \mapsto T_t)
\;.
$$
This holds because $O_1T_1O_1^*=O_1A^*O_0^*(O_0T_0O_0^*)O_0AO_1^*$ and $\det(O_0AO_1^*)=\det(A)$ since $O_1$ and $O_0$ are in the same component of the orthogonal group. This transposes to an invariance property for the parity. Similarly, other properties of parity result from properties of the $\ZM_2$-valued spectral flow.

\vspace{.2cm}

Let us next provide some examples that illustrate the topological stability associated to the parity. For $N=1$ we first consider two paths
\begin{equation}
\label{eq-examp}
T_t\;=\;\begin{pmatrix} 0 & t \\ -t & 0 \end{pmatrix}
\;,
\qquad
\widetilde{T}_t\;=\;\begin{pmatrix} 0 & |t| \\ -|t| & 0 \end{pmatrix}
\;,
\qquad
t\in[-1,1]
\;.
\end{equation}
Clearly these two paths are isospectral $\spec(T_t)=\spec(\widetilde{T}_t)$ for all $t \in [-1,1]$. By \eqref{eq-PFfinite2} one finds $\SF_2([-1,1]\ni t\mapsto T_t)=-1$ and $\SF_2([-1,1]\ni t \mapsto \widetilde{T}_t)=1$. This has spectral consequences. The latter can be perturbed to $\widetilde{T}_t(s)$ (within the class of real chiral skew-adjoints)  in such a way that $0$ is not an eigenvalue for any $t$:
$$
\widetilde{T}_t(s)\;=\;\begin{pmatrix} 0 & |t|+s \\ -(|t|+s) & 0 \end{pmatrix}
\;.
$$
Indeed, the eigenvalues are then $\pm \imath(|t|+s)$ which both never vanish for positive $s>0$. It is not possible to construct such a perturbation for $T_t$, namely any real skew-adjoint perturbation conserving the chiral symmetry can merely shift the eigenvalue crossing at $0$. 

\vspace{.2cm}

Furthermore, let us double the non-trivial example in \eqref{eq-examp} via a direct sum to $T'_t=T_t\oplus T_t$ which is chiral with respect to $J \oplus J=\diag(1,-1,1,-1)$. Now $T'_t$ is block diagonal rather than in the off-diagonal form \eqref{eq-link}. However, using the permutation $U$ of the second and third component one obtains the spectral representation $U (J\oplus J) U^*=\diag(\one,-\one)$ and then $UT'_tU^*$ is of the form \eqref{eq-link} with off-diagonal entry $B'_t=\diag(t,t)$. Then by the multiplicativity of the $\ZM_2$-valued spectral flow, $\SF_2([-1,1]\ni t\mapsto T'_t)=(-1)(-1)=1$. Again it is then possible to lift the kernel along the whole path by a real chiral skew-adjoint perturbation. One such perturbation is 
$$
U T'_t(s) U^*
\;=\;
\begin{pmatrix}
0 & 0 & t & -s \\
0 & 0 & s & t \\
-t & -s & 0 & 0 \\
s & -t & 0 & 0
\end{pmatrix}
\;.
$$
Indeed, the spectrum of $T'_t(s)$ is $\{\imath(t^2+s^2)^{\frac{1}{2}},-\imath(t^2+s^2)^{\frac{1}{2}}\}$ with a double degeneracy. In particular, for $s\not=0$, $T'_t(s)$ is invertible for all $t\in[-1,1]$.

\section{Construction of the parity in infinite dimension}
\label{sec-InfDim}

In this section, the separable real Hilbert space $\Hh_\RM$ is now of infinite dimension and the continuous path $[0,1]\ni t\mapsto B_t\in\Ff^0$ is within the Fredholm operators. For the sake of simplicity, let us first suppose that it lies in the component of Fredholm operators with vanishing index. (In a large part of the literature these are called Fredholm indices even though it was actually F. Noether who first exhibited a Fredholm operator with non-vanishing index \cite{Die}.) The general case will then be dealt with towards the end of the section. In \cite{FP0,FP}, the parity of an admissible path (namely with invertible endpoints) uses the Leray-Schauder degree which is defined as follows. One first proves that there exists a second path of real invertibles $[0,1]\ni t\mapsto M_t$ such that $M_tB_t=\one+K_t$ with a real compact operator $K_t$. Then, if $n_t$ denotes the number of negative eigenvalues of $\one+K_t$ counted with multiplicity, $n_t$ coincides with the number of eigenvalues less than $-1$ of the compact operator $K_t$ and is therefore finite. The (linear) Leray-Schauder degree is 
\begin{equation}
\label{eq-DetDeg}
\degLS(B_t)\;=\;(-1)^{n_t}\in\ZM_2
\;.
\end{equation}
Let us explain how this fits together with Definition~\ref{def-PFfinite}. If $B_t$ is a matrix, one can choose $M_t=\one$; the spectrum of $B_t$ is symmetric with respect to the reflection on the real axis; now non-real eigenvalues of $B_t$ come in complex conjugate pairs which do not contribute to $\sgn(\det(B_t))$; hence analyzing the real eigenvalues immediately leads to $\degLS(B_t)=\sgn(\det(B_t))$. For the path, the parity is then as in Definition~\ref{def-PFfinite} given by $\PF([0,1]\ni t\mapsto B_t)=\degLS(B_1)\degLS(B_0)\in\ZM_2$ \cite{FP}. One of the difficulties with this approach is that, in general, it is very hard to determine the path $M_t$ and therefore also the parity by this procedure.

\vspace{.2cm}

This work provides an alternative approach in which the parity is defined as the $\ZM_2$-valued spectral flow studied in \cite{CPS} of a path of skew-adjoint operators on the doubled Hilbert space. This is based on the passage \eqref{eq-link} to chiral skew-adjoint operators. Hence let $\Hh'_\RM=\Hh_\RM\oplus\Hh_\RM$ be a real Hilbert space equipped with the $\ZM_2$-grading $J=\diag(\one,-\one)$. Then \eqref{eq-link} identifies $\Ff^0=\Ff^0(\Hh_\RM)$ with
$$
\hat{\Ff}^0
\;=\;
\left\{T\in\Bb(\Hh'_\RM)\,:\,T=-T^*=-JTJ\;\;\mbox{\rm Fredholm}
\right\}
\;.
$$
Hence $\hat{\Ff}^0$ is a subspace of $\Ff^1=\{T\in\Bb(\Hh'_\RM)\,:\,T=-T^*\;\mbox{\rm Fredholm}\}$, and any path in $\Ff^0$ can be viewed as a path $[0,1]\ni t\mapsto T_t\in \hat{\Ff}^0$ in $\Ff^1$. This path has a supplementary chiral symmetry $JT_tJ=-T_t$, but this is irrelevant for the definition of its $\ZM_2$-valued spectral flow that we review next. Hence let now $[0,1]\ni t\mapsto T_t\in {\Ff}^1$. As already mentioned, we will first deal with an admissible path with invertible endpoints $T_0$ and $T_1$. Roughly, the idea is to reduce the definition of the $\ZM_2$-valued spectral flow to the finite dimensional definition by extracting from $T_t\in\Ff^1$ only the finite-dimensional subspace corresponding to eigenvalues in a small interval around $0$, just as in \cite{Ph}. Thus, for $a>0$ let us set
$$
Q_{a}(t)\;=\;
\chi_{(-a,a)}(\imath\,T_t)
\;,
$$
where $\chi_I$ denotes the characteristic function on $I\subset\RM$. The projection $Q_{a}(t)$ is of finite dimensional range for $a$ sufficiently small by the Fredholm property of $T_t$. Associated to these projections, one has the restrictions $Q_{a}(t)\,T_t\,Q_{a}(t)$ which are viewed as skew-adjoint matrices on $\Ee_a(t)=\Ran(Q_{a}(t))$. By compactness (see the first Lemma in \cite{Ph}), it is possible to choose a finite partition $0=t_0<t_1<\ldots<t_{N-1}<t_N=1$ of $[0,1]$ and $a_n> 0$, $n=1,\ldots,N$, such that each piece $ [t_{n-1},t_n]\ni t\mapsto Q_{a_n}(t)$ is continuous and hence of constant finite rank, and, moreover, for some $\epsilon$,
\begin{equation}
\label{eq-Qcont}
\|Q_{a_n}(t)-Q_{a_n}(t')\|\;<\;\epsilon
\;,
\qquad
\forall\;\;t,t'\in[t_{n-1},t_n]
\;.
\end{equation}
Let $\isom_n:\Ee_{a_n}(t_{n-1})\to \Ee_{a_n}(t_{n})$ be the orthogonal projection of $\Ee_{a_n}(t_{n-1})$ onto $\Ee_{a_n}(t_{n})$, namely $\isom_nv=Q_{a_n}(t_{n})v$. Then $\isom_n$ is a bijection allowing to identify $\Ee_{a_n}(t_{n-1})$ with $\Ee_{a_n}(t_{n})$. Now each interval $ [t_{n-1},t_n]$ leads to a path $ [t_{n-1},t_n]\ni t\mapsto Q_{a_n}(t)\,T_t\,Q_{a_n}(t)$ of chiral, skew-adjoint matrices on $\Ee_{a_n}(t)=\Ran(Q_{a_n}(t))$, but this path may not be admissible. To lift the (even-dimensional) kernel at the endpoint $t_n$, one can add a skew-adjoint perturbation $R_{n}$ on the kernel of $Q_{a_n}(t)\,T_{t_n}\,Q_{a_n}(t_n)$ so that 
\begin{equation}
\label{eq-Tachoice}
T^{(a_n)}_{t_n}\;=\;Q_{a_n}(t_n)\,T_{t_n}\,Q_{a_n}(t_n)\,+\,R_n
\;,
\end{equation}
are skew-adjoint invertible operators on $\Ee_a(t)$. Clearly the choice of the $R_n$ is largely arbitrary, but it is part of Theorem~\ref{theo-welldef} below that the following definition is independent of the choice of the $R_n$.

\begin{defini}
\label{def-Z2flow}
For an admissable path $[0,1]\ni t\mapsto T_t\in \Ff^1$, let $t_n$ and $a_n$ as well as $T^{(a)}_t$ and $\isom_n$ be as above. Then the  $\ZM_2$-valued spectral flow is defined by 
\begin{equation}
\label{eq-SFdef}
\SF_2([0,1]\ni t\mapsto T_t)
\;=\;
\prod_{n=1,\ldots,N}
\SF_2\big(T^{(a_n)}_{t_{n-1}},\isom_n^*T^{(a_n)}_{t_n}\isom_n\big)
\;,
\end{equation}
where on the right hand side the $\SF_2$ is the finite dimensional $\ZM_2$-valued spectral flow on $\Ee_{a_n}(t_{n-1})$ as given in \eqref{eq-PFfinite2}, and the product is in the multiplicative group $(\ZM_2,\cdot)$.
\end{defini}

Let us stress that in infinite dimension, it is in general {\it not} possible to write $\SF_2(T_0,T_1)$ for $\SF_2([0,1]\ni t \mapsto T_t)$ because the $\ZM_2$-valued spectral flow depends on the choice of the path. The basic result on the $\ZM_2$-valued spectral flow is that it is well-defined by the above procedure. 

\begin{theo}[Theorem~4.2 in \cite{CPS}]
\label{theo-welldef} 
Let $[0,1]\ni t\mapsto T_t\in\Ff^1$ be an admissible path. The definition of $\SF_2([0,1] \ni t\mapsto T_t)$ is independent of the choice of the partition $0=t_0<t_1<\ldots<t_{N-1}<t_N=1$ of $[0,1]$ and the values $a_n>0 $ such that $ [t_{n-1},t_n]\ni t \mapsto Q_{a_n}(t)$ is continuous and satisfies \eqref{eq-Qcont}, and also the choice of the $R_n$ in \eqref{eq-Tachoice}. 
\end{theo}

As $\hat{\Ff}^0\subset \Ff^1$, one can now use the $\ZM_2$-valued spectral flow to define the parity.

\begin{defini}
\label{def-PFflow}
Let $[0,1]\ni t\mapsto B_t\in\Ff^0$ be an admissible path and $[0,1]\ni t\mapsto T_t\in \hat{\Ff}^0$ be the path associated by \eqref{eq-link}. Then the parity is defined by 
\begin{equation}
\label{eq-PFlink}
\sigma([0,1]\ni t\mapsto B_t)
\;=\;
\SF_2([0,1]\ni t\mapsto T_t)\;\in\;\ZM_2
\;.
\end{equation}
\end{defini}

Let us stress again that $T_t\in\hat{\Ff}^0$ implies the chiral symmetry $JT_tJ=-T_t$, but this is not of importance for the definition of the $\ZM_2$-valued spectral flow on the right-hand side of \eqref{eq-PFlink}. One can, however, make more specific choices in the construction of $\SF_2$ above, notably the spectral projections satisfy due to the symmetry of $[-a,a]$
$$
Q_{a}(t)\;=\;JQ_a(t)J
\;,
$$
and one can choose the skew-adjoint perturbations $R_{n}$ to be chiral. Due to Definition~\ref{def-PFflow}, the parity inherits from the $\ZM_2$-valued spectral flow all of the properties stated in \cite{CPS}. They are collected in the following result. Most of these properties are already stated in Chapter~6 of \cite{FP}. 

\begin{theo}
\label{theo-Properties} 
Let $[0,1]\ni t \mapsto B_t\in\Ff^0$ be an admissible path. 

\begin{enumerate}

\item[{\rm (i)}] The parity is homotopy invariant under homotopies in the paths of  Fredholm operators keeping the endpoints fixed.

\item[{\rm (ii)}] If $B_t$ is invertible for all $t\in[0,1]$, then $\PF([0,1]\ni t\mapsto B_t)=1$.

\item[{\rm (iii)}] The parity has a concatenation property, namely if $[0,2]\ni t\mapsto B_t\in\Ff^0$ is a path such that $B_2$ is invertible, then
$$
\PF([0,1]\ni t\mapsto B_t)
\;\cdot\;
\PF([1,2]\ni t\mapsto B_{t})
\;=\;
\PF([0,2]\ni t\mapsto B_{t})
\;.
$$ 

\item[{\rm (iv)}] The parity is independent of the orientation of the path:
$$
\PF([0,1]\ni t\mapsto B_t)
\;=\;
\PF([0,1]\ni t\mapsto B_{1-t})
\;.
$$ 

\item[{\rm (v)}] The parity has a multiplicativity property under direct sums, namely if $[0,1]\ni t\mapsto B'_t\in\Ff^0$ is a second admissible path,
$$
\PF([0,1]\ni t\mapsto B_t\oplus B'_t )
\;=\;
\PF([0,1]\ni t\mapsto B_t)\,\cdot\,\PF([0,1]\ni t\mapsto B'_t )
\;.
$$

\item[{\rm (vi)}] The parity is invariant under the conjugation by a path $[0,1]\ni t \mapsto O_t$ of orthogonals:
$$
\PF([0,1]\ni t \mapsto O_tB_tO_t^*)
\;=\;
\PF([0,1]\ni t \mapsto B_t)
\;.
$$
In particular, the parity is independent under reflection of the path:
$$
\PF([0,1]\ni t\mapsto B_t)
\;=\;
\PF([0,1]\ni t\mapsto -B_{t})
\;.
$$ 

\end{enumerate}

\end{theo}

The following result is already stated in \cite{FP}. 

\begin{theo} 
\label{theo-isomorphism}
The map $\PF$ on loops in $\Ff^0$ is a homotopy invariant and induces an isomorphism of $\pi_1(\Ff^0)$ with $\ZM_2$.
\end{theo}

\noindent {\bf Proof.}  As $\pi_1(\Ff^0)\cong\ZM_2$ is already known \cite{AS} and $\PF$ is homotopy invariant, one only has to check that $\PF$ takes two different values on the two different components of the based loop space in $\Ff^0$. For constant paths (and thus all contractible ones) the parity vanishes. An example with a parity equal to $-1$ is given in Section~\ref{sec-example}.
\hfill $\Box$

\vspace{.2cm}

Up to now, only paths $[0,1]\ni t\mapsto B_t$ in the component of $\Ff^0$ with vanishing index were considered. For general paths, one has
$$
\dim\big(\Ker(T_t)\big)\, -\,\big|\Ind(B_t)\big|
\;\in\;
2\,\NM_0
\;.
$$
In particular, for non-vanishing $\Ind(B_t)$ the dimension of the kernel of $T_t$ is positive for all $t$ and thus there are no admissible paths. However, there are several possibilities to reduce this case to the prior one. For that purpose, let us now call a path admissible if $\dim\big(\Ker(T_i)\big)=\big|\Ind(B_i)\big|$ for $i=0,1$. Recall that $\Ker(T_t)$ is $J$-invariant. Let now $[0,1]\ni t\mapsto P_t$ be a continuous path of $J$-invariant orthogonal projections onto parts of the kernel of $T_t$, and being of the dimension of $\Ker(T_i)$ for $i=0,1$. Then follow the constructions and arguments from above for $T_t$ restricted to the range of $\one-P_t$. This construction is independent of the choice of $[0,1]\ni t\mapsto P_t$ for, if $[0,1]\ni t\mapsto P'_t$ is another projection with the above properties, then $P'_t=O_tP_t O_t^*$ for a path $[0,1]\ni t\mapsto O_t$ of orthogonals commuting with $J$. Hence it follows from property (vi) of Theorem~\ref{theo-Properties} that one obtains the same parity. All properties of Theorem~\ref{theo-Properties} transpose directly, except for (ii) which now states that paths with constant nullity have a parity equal to $1$.

\section{Reformulation with chiral self-adjoints} 
\label{sec-Extensions}

Given an admissible path $[0,1]\ni t \mapsto B_t$ of Fredholm operators on $\Hh_\RM$,  it is possible to associate self-adjoint real operators on $\Hh'_\RM=\Hh_\RM\oplus\Hh_\RM$ via
\begin{equation}
\label{eq-link2}
H_t
\;=\;
\begin{pmatrix}
0 & B_t\\
B_t^* & 0
\end{pmatrix}
\;.
\end{equation}
This identifies $\Ff^0$ with the set $\tilde{\Ff}^0$ of chiral self-adjoint Fredholm operators
$$
\tilde{\Ff}^0
\;=\;
\left\{H\in\Bb(\Hh'_\RM)\,:\,H=H^*=-JHJ\;\;\mbox{\rm Fredholm}
\right\}
\;.
$$
A bijection between $\hat{\Ff}^0$ and $\tilde{\Ff}^0$ is given by
$$
\hat{\Ff}^0
\;=\;
\imath\,J^\frac{1}{2} \,
\tilde{\Ff}^0\,(J^\frac{1}{2} )^*
\;,
$$
where $J^\frac{1}{2}=\diag(\one, \imath \one)$ is the square root of $J$. In some applications (as in Section~\ref{sec-TopIns}) one rather finds admissible paths $[0,1]\ni t \mapsto H_t\in\tilde{\Ff}^0$ of chiral self-adjoint real operators. Such paths then have a parity given by
$$
\PF([0,1]\ni t\mapsto H_t)
\;=\;
\SF_2\big([0,1]\ni t\mapsto \imath\,J^\frac{1}{2} \,H_t\,(J^\frac{1}{2} )^*\big)
\;.
$$

A further modification concerns a setting with complex Hilbert spaces and a reality condition involving another symmetry. Suppose thus that one has a complex Hilbert space $\Hh_\CM$ with a real structure given by a (anti-linear involutive) complex conjugation $\Cc:\Hh_\CM\to\Hh_\CM$, naturally extended to $\Hh'_\CM=\Hh_\CM\oplus\Hh_\CM$. For any linear operator $A$ on $\Hh_\CM$ or $\Hh'_\CM$ let us set $\overline{A}=\Cc A\Cc$. Further suppose given a real self-adjoint involution $K$ on $\Hh'_\CM$, namely $\overline{K}=K^*=K$ and $K^2=\one$ which, moreover, commutes with $J$. Then an operator $A$ is called $K$-real if $K^*\overline{A}K=A$. Now one considers admissible paths $[0,1]\ni t\mapsto H_t$ of $K$-real self-adjoint chiral operators, namely
$$
K^*\overline{H_t}K\;=\;H_t
\;,
\qquad
H_t^*\;=\;H_t
\;,
\qquad
J^*{H_t}J\;=\;-H_t
\;.
$$
Also for such paths one can define the parity. Indeed, let $L$ be the root of $K$ with spectrum $\{1,\imath\}$. It commutes with $J$. Then set
$$
\widehat{H}_t\;=\;L^*H_tL\;.
$$
It can be checked that $\widehat{H}_t$ is real, self-adjoint and chiral with respect to $J$. Consequently, $\widehat{H}_t$ can be restricted to an $\RM$-linear operator on $\Hh'_\RM=\Ker(\Cc-\one)\subset\Hh'_\CM$. Thus it is within the class of paths considered above and the parity of $[0,1]\ni t\mapsto H_t$ can be defined as that of $[0,1]\ni t\mapsto \widehat{H}_t$.

\section{Fredholm pairs of chiral complex structures} 
\label{sec-PI}

The aim of this section is to construct an alternative formula for the parity. This will first be done for special paths between complex structures that are close in the Calkin algebra, then later on it will also be extended to general paths. Recall that a complex structure $I$ on $\Hh'_\RM$ is a linear, skew-adjoint and unitary operator on $\Hh'_\RM$. It is called chiral if, moreover, $JIJ=-I$ for a symmetry $J$, namely $J=J^*=J^{-1}$. Hence any chiral complex structure is an element of $\hat{\Ff}^0$. The following definition is motivated by \cite{CO,CPS}, as well as Kato's Fredholm pair of projections and its index \cite{Kat,ASS}.

\begin{defini} 
\label{def-FredPair}
A pair $(I_0,I_1)$ of chiral complex structures on $\Hh'_\RM$ is called a Fredholm pair of chiral complex structures if $\|\pi(I_0-I_1)\|_\Qq<2$. The $\ZM_2$-index of $(I_0,I_1)$ is then defined by
\begin{equation}
\PI(I_0,I_1)
\;=\; 
\Big(
\tfrac{1}{2}\,\dim_\RM(\Ker_\RM(I_0+I_1))\Big)
\;\mbox{\rm mod}\;2\;\in\;\ZM_2
\;.
\label{eq-jdef}
\end{equation}
\end{defini}

The index is indeed well-defined because $I_0+I_1=2\,I_0+(I_1-I_0)$ has no essential spectrum at $0$ and it will be shown in the proof of Theorem~\ref{theo-jproperties} that the kernel of $I_0+I_1$ is even dimensional. On the right hand side of \eqref{eq-jdef} the additive version of $\ZM_2$ was used and is tacitly identified with the multiplicative one. The following justifies Definition~\ref{def-FredPair}.

\begin{theo} 
\label{theo-jproperties}
The map $(I_0,I_1)\mapsto \PI(I_0,I_1)\in \ZM_2$ is a homotopy invariant on the set of Fredholm pairs of chiral symmetries. Moreover, for both signs one has
\begin{equation}
\PI(I_0,I_1)
\;=\; 
\dim_\RM(\Ker_\RM(I_0-I_1\pm\,2\,\imath\,\one))
\;\mbox{\rm mod}\;2
\;.
\label{eq-jdef2}
\end{equation}
\end{theo}

Before going into the proof, let us elaborate on the connection to the index of a Fredholm pair of projections \cite{ASS}. Here there are two projections $P_0=\frac{1}{2}(\imath I_0+\one)$ and  $P_1=\frac{1}{2}(\imath I_1+\one)$ associated to the complex structures. The property $\|\pi(I_0-I_1)\|_\Qq<2$ is equivalent to $\|\pi(P_0-P_1)\|_\Qq<1$ and thus to $(P_0,P_1)$ being a Fredholm pair. Furthermore, these two projections satisfy $\overline{P_j}=\one-P_j$ and $J\overline{P_j}J= P_j$. In the terminology of \cite{GS} this means that the $P_j$ are even real and even Lagrangian projections. These symmetries imply that the index of the Fredholm pair $(P_0,P_1)$ vanishes, namely the two signs on the right hand side of \eqref{eq-jdef2} lead to the same dimension (compare with eq. (3.1) in \cite{ASS}). Hence one sees that the $\ZM_2$-index $\PI(I_0,I_1)$ is a secondary invariant associated to the Fredholm pair $(P_0,P_1)$ which is well-defined due to Theorem~\ref{theo-jproperties}. 


\vspace{.2cm}

The proof of Theorem~\ref{theo-jproperties} will be based on the following lemma in which the chiral symmetry and reality are irrelevant. The lemma can be traced back to \cite{CO} and is stated as Lemma~5.3 in \cite{CPS}. An equivalent algebraic fact has also been used for pairs of orthogonal projections \cite[Theorem 2.1]{ASS}.

\begin{lemma}
\label{lem-Ident} 
Let $I_0$ and $I_1$ be complex structures. Set 
$$
T_0\;=\;\tfrac{1}{2}\,(I_0+I_1)
\;,
\qquad
T_1\;=\;\tfrac{1}{2}\,(I_0-I_1)
\;.
$$
Then the following identities hold:
$$
T_0^*T_0\,+\,T_1^*T_1\;=\;\one\;=\;T_0T_0^*+T_1T_1^*\;,
\qquad
T_0^*T_1+T_1^*T_0\;=\;0\;=\;T_0T_1^*+T_1T_0^*
\;,
$$
as well as
$$
T_0I_0\;=\;I_1T_0\;,
\qquad
T_0I_1\;=\;I_0T_0\;,
\qquad
T_1I_0\;=\;-I_1T_1\;,
\qquad
T_1I_1\;=\;-I_0T_1\;.
$$
\end{lemma}

\noindent {\bf Proof.} Everything is verified by straightforward computations. 
\hfill $\Box$

\vspace{.2cm}

\noindent {\bf Proof} of Theorem~\ref{theo-jproperties}. First of all, as noted above $T_0=I_0+\frac{1}{2}(I_1-I_0)$ is a skew-adjoint Fredholm operator by the assumption that $(I_0,I_1)$ is a Fredholm pair so that there is only discrete spectrum in a neighborhood of $0$. The idea of the proof is to show that every (small) eigenvalue of $T_0$ of finite multiplicity has even multiplicity. As $T_0$ is chiral and its spectrum satisfies $\spec(T_0)=-\spec(T_0)$, this then implies that the nullity of $T_0$ only changes by multiples of $4$ under homotopic changes of $T_0$ (induced by a homotopy of $I_0$ and $I_1$). The main tool is to view $I_0$ as a complex structure on $\Hh'_\RM$.  Now $(T_0^*T_0) I_0=-T_0I_1T_0=I_0(T_0^*T_0)$ so that $T_0^*T_0$ is a complex linear operator on $\Hh'_\RM$ viewed as complex Hilbert space (using the complex structure $I_0$). Consequently the real multiplicity of all eigenvalues of $T_0^*T_0$ is even. In particular, $\PI (I_0,I_1)$ given by \eqref{eq-jdef} indeed takes values in $\{0,1\}$. Next let us show that for $\lambda\in(0,1)$, the complex multiplicity of the eigenspaces of $T_0^*T_0$ is a multiple of $2$ (then the  real multiplicity is a multiple of $4$). Suppose that $T_0^*T_0v=\lambda v$ for some non-vanishing vector $v$. Then set $w=T_1^*T_0v$. First of all, its norm does not vanish:
$$
\|w\|^2\;=\;v^*T_0^*T_1T_1^*T_0v\;=\;v^*T_0^*(\one-T_0T_0^*)T_0v\;=\;\lambda(1-\lambda)\|v\|^2
\;.
$$
It is also an eigenvector of $T_0^*T_0$:
$$
T_0^*T_0w\;=\;T_0^*T_0T_1^*T_0v\;=\;-T_0^*T_1T_0^*T_0v\;=\;T_1^*T_0T_0^*T_0v\;=\;\lambda w
\;.
$$
Moreover, it is complex linearly independent of $v$. In fact, suppose the contrary, namely that $w=(\mu_0+\mu_1I_0)v$ for some $\mu_0,\mu_1\in\RM$. Multiplying this with $T_0^*T_1$ leads to
$$
\lambda(1-\lambda)v\;=\;T_0^*T_1T_1^*T_0v\;=\;T_0^*T_1w\;=\;T_0^*T_1(\mu_0+\mu_1I_0)v
\;=\;
-(\mu_0-\mu_1I_0)w
\;,
$$
where in the last equality the identity $T_0^*T_1I_0=-I_0T_0^*T_1$ was used. Multiplying now by $(\mu_0+\mu_1I_0)$ shows
$$
\lambda(1-\lambda)w\;=\;-(\mu_0^2+\mu_1^2)w
\;,
$$
that is, a contradiction. If there are further eigenvectors of $T_0^*T_0$ with eigenvalue $\lambda$, one can restrict to the orthogonal complement and iterate the above argument.

\vspace{.1cm}

As to the alternative formula for $\PI (I_0,I_1)$, let us note that the kernel of $T_0^*T_0$ coincides with the eigenspace of $T_1^*T_1$ to the eigenvalue $1$, which in turn is given by the direct sum of the eigenspaces of $T_1$ for the eigenvalues $\imath$ and $-\imath$. This proves the formula. Let us comment that another proof of the homotopy invariance uses the chiral symmetry of $T_1$ and checks the double degeneracy of all eigenvalues of $T_1$ in $(0,1)$, excluding $1$. 
\hfill $\Box$

\vspace{.2cm}

The following result establishes the link of $\PI (I_0,I_1)$ with the $\ZM_2$-valued spectral flow of the straight line connecting $I_0$ and $I_1$, which indeed lies in $\hat{\Ff}^0\subset\Ff^1$. 

\begin{proposi} 
\label{theo-Pf2j}
For any Fredholm pair of chiral complex structures $(I_0,I_1)$ on $\Hh'_\RM$, one has
$$
\SF_2\big([0,1]\ni t\mapsto (1-t)\,I_0\,+\,t\,I_1\big)
\;=\;
\PI (I_0,I_1)
\;.
$$

\end{proposi}

\noindent {\bf Proof.} This is essentially identical to the argument leading to Proposition~6.2 in \cite{CPS}, so let us just give a sketch. The operators $T_t=(1-t)I_0+t I_1$ are indeed Fredholm because, for $t\in[0,\frac{1}{2}]$, $T_t=I_0+t(I_1-I_0)$ is a perturbation of an operator $I_0$ with spectrum $\{-\imath,\imath\}$ by an operator with bound $1$ in the Calkin algebra so that $T_t$ has its essential spectrum bounded away from $0$. For $t\in[\frac{1}{2},1]$, this holds by the same argument as $T_t=I_1+(1-t)(I_0-I_1)$. Moreover, $T_t$ is  invertible except possibly at $t=\frac{1}{2}$. Hence in Definition~\ref{def-Z2flow} it is sufficient to work with three intervals $[0,\frac{1}{2}-\epsilon]$, $[\frac{1}{2}-\epsilon,\frac{1}{2}+\epsilon]$ and $[\frac{1}{2}+\epsilon,1]$ for some $\epsilon>0$. Only the middle interval has a possibly non-vanishing contribution coming from the parity of the nullity of $T_{\frac{1}{2}}=\frac{1}{2}(I_0+I_1)$. But this is precisely the definition \eqref{eq-jdef} of the $\ZM_2$-index.
\hfill $\Box$

\vspace{.2cm}

Further following \cite{Ph1} or \cite{CPS}, one can go on and rewrite the definition of the parity. 

\begin{proposi}
\label{prop-AltForm}
Let $[0,1]\ni t\mapsto B_t\in{\Ff}^0$ be an admissible path and associated $T_t\in\hat{\Ff}^0$ to $B_t$ by  \eqref{eq-link}. Let $I_t$ be chiral complex structures obtained by completing the phase $T_t|T_t|^{-1}$ on the kernel. Then, for a sufficiently fine partition $0=t_0<t_1\cdots< t_N=1$ satisfying $\|\pi(I_n-I_{n-1})\|<2$, one has for the parity
$$
\sigma([0,1]\ni t\mapsto B_t)
\;=\;
\Big(
\sum_{n=1,\ldots,N} \PI (I_{t_{n-1}},I_{t_{n}})
\Big)
\;\mbox{\rm mod}\;2
\;.
$$
\end{proposi}

\noindent {\bf Proof.} Let us begin by rewriting Definition~\ref{def-Z2flow}. One can choose $R_n$ in \eqref{eq-Tachoice} sufficiently small and the partition $t_0=0<t_1<\cdots<t_N=1$ sufficient fine such that $a_n$ from Definition~\ref{def-Z2flow} is not in the spectrum of $(1-t)\,\imath \,(T_{t_{n-1}}+R_{n-1})+t\,\imath\,(T_{t_n}+R_{n})$ for any $t \in [0, 1]$. By definition 
$$
\SF_2\big(T^{(a_n)}_{t_{n-1}},\isom_n^*T^{(a_n)}_{t_n}\isom_n\big) 
\; = \;
 \SF_2\big([0,1]\ni t\mapsto (1-t)\,(T_{t_{n-1}}+R_{n-1})+t\,(T_{t_n}+R_{n})\big)
$$
and
$$
\SF_2([0,1]\ni t\mapsto T_t)
\;=\;
\prod_{n=1,\ldots,N} 
\SF_2\big([0,1]\ni t\mapsto (1-t)\,(T_{t_{n-1}}+R_{n-1})+t\,(T_{t_n}+R_{n})\big)
\;.
$$
By identifying $T_{t_n}+R_{n}$ with $T_{t_n}$, we can from now on assume that $T_{t_n}$ is invertible. Next let us claim that for each $n=1,\ldots,N$ one has
$$
\label{eq-PFPI}
\SF_2\big([0,1]\ni t\mapsto (1-t)\,T_{t_{n-1}}+t\,T_{t_n}\big)
\;=\;
\SF_2\big([0,1]\ni t\mapsto (1-t)\,I_{t_{n-1}}+t\,I_{t_n}\big)
\;.
$$
Indeed, as $T_{t_n}$ and $T_{t_{n-1}}$ are both invertible, 
$$
[0,1]\ni s\;\mapsto\; (1-t)\,T_{t_{n-1}}|T_{t_{n-1}}|^{-s}+t\,T_{t_n}|T_{t_n}|^{-s}
$$ 
deforms the initial path into the path $[0,1]\ni t\mapsto (1-t)\,I_{t_{n-1}}+t\,I_{t_n}$. During this homotopy the endpoints remain invertible so that the $\ZM_2$-valued spectral flow is unchanged. Now the assertion follows from Proposition~\ref{theo-Pf2j}.
\hfill $\Box$

\section{Parity of paths between unitary conjugates}
\label{sec-index}

Let $\Hh'_\RM=\Hh_\RM\oplus\Hh_\RM$ be equipped with the $\ZM_2$-grading $J=\diag(\one,-\one)$. The orthogonal group preserving $J$ is
$$
\Oo(\Hh'_\RM,J)
\;=\;
\{
O\in\Oo(\Hh'_\RM)\;:\;
O^*JO=J
\}
\;.
$$ 
This is a subgroup of $\Oo(\Hh'_\RM)$ naturally identified with $\Oo(\Hh_\RM)\times \Oo(\Hh_\RM)$ because $O^*JO=J$ is equivalent to $JOJ=O$ which requires $O$ to be diagonal in the grading of $J$.  For any real chiral complex structure $I$, let us set
\begin{equation} 
\label{eq-OQdef}
\Oo_I(\Hh'_\RM,J)
\;=\;
\{
O\in \Oo(\Hh'_\RM,J)
\;:\;
[O,I]\in\Kk(\Hh'_\RM)
\}
\;,
\end{equation}
where $\Kk(\Hh'_\RM)$ denotes the compact operators on $\Hh'_\RM$. This is a subgroup of $\Oo(\Hh'_\RM,J)$. Let us note that for $O\in \Oo_I(\Hh'_\RM,J)$ one has $\pi(O^*IO)=\pi(I)$ in the Calkin algebra. Furthermore, recall the definition of the based loop space $\Omega_I\hat{\Ff}^0$ of $\hat{\Ff}^0$ based at $I$:
$$
\Omega_I\hat{\Ff}^0\;=\;
\big\{[0,1]\ni t \mapsto T_t \in \hat{\Ff}^0\; :\; T_1=T_0=I\big\}
\;.
$$

\begin{theo} 
\label{theo-IsoQqQ} 
For any chiral complex structure $I$ on $\Hh'_\RM$, the group $\Oo_I(\Hh'_\RM,J)$ is homotopy equivalent to $\Omega_I\hat{\Ff}^0$. In particular,
$\pi_0(\Oo_I(\Hh'_\RM,J))\cong\ZM_2$.
\end{theo}

As a preparatory result for the proof, let us state the following.

\begin{proposi} 
\label{prop-IsoF_0} 
The space $\Ff^0\cong \hat{\Ff}^0$ is homotopy equivalent to the space
$$
\Cc(\Hh'_\RM)
\;=\;
\{\pi(I)\in\Qq\,:\,\pi(I)\;\mbox{\rm chiral complex structure}\}
\;.
$$
\end{proposi}

\noindent {\bf Proof.} We closely follow the proof of Theorem~7.1 of \cite{CPS}, which in turn is based on \cite{AS,Ph1}.
Let $\rho:\hat{\Ff}^0\to\hat{\Ff}^0$ be the (non-linear and discontinuous) map sending $T$ to the partial isometry $I=T|T|^{-1}$ in the polar decomposition. If $\pi$ denotes as before the quotient map onto the Calkin algebra $\Qq=\Qq(\Hh'_\RM)$ over $\Hh'_\RM$, then the map $\hat{\rho}=\pi\circ\rho$ sends $\hat{\Ff}^0$ surjectively onto $\Cc(\Hh'_\RM)$. Indeed, any chiral complex structure $\pi(I)\in\Cc(\Hh'_\RM)$ has a chiral and skew-adjoint lift $I'$ for which $(I')^*I'-\one$ is compact; then the Riez projections $P'_\pm$ on the positive and negative spectral projections of $-\imath I'$ lead to a lift $I=\imath P'_+-\imath P'_-$ for which $I^*I-\one$ is a finite dimensional projection (on the kernel of $I'$). The Bartle-Graves selection theorem \cite{BD} now provides a right inverse $\theta:\Cc(\Hh'_\RM)\to \hat{\Ff}^0$ to $\hat{\rho}$, namely $\hat{\rho}\circ \theta=\one$. Moreover, $\theta\circ \hat{\rho}$ is homotopic to the identity via $[0,1]\ni t\mapsto t\,T+(1-t)\,\theta(\hat{\rho}(T))$. As $\theta(\hat{\rho}(T))=T|T|^{-1}+K$ for some chiral skew-adjoint compact $K$, this is a homotopy in $\hat{\Ff}^0$. Thus $\hat{\rho}$ is actually a homotopy equivalence so that $\hat{\Ff}^0$ and $\Cc(\Hh'_\RM)$ are homotopy equivalent.
\hfill $\Box$

\vspace{.2cm}

\noindent {\bf Proof} of Theorem~\ref{theo-IsoQqQ}. Due to Proposition~\ref{prop-IsoF_0} it is sufficient to show the homotopy equivalence of $\Oo_I(\Hh'_\RM,J)$ and $\Omega_{\hat{\rho}(I)}\Cc(\Hh'_\RM)$. Here, the chiral complex structure $I$ on $\Hh'_\RM$ also specifies a base point $\hat{\rho}(I)$ in $\Cc(\Hh'_\RM)$. Associated to $I$, one can define a map $\beta_I:\Oo(\Hh'_\RM,J)\to\Cc(\Hh'_\RM)$ via $\beta_I(O)=\hat{\rho}(OIO^*)=\pi(OIO^*)$. This map is actually a Serre fibration by the argument in Theorem 3.9 of \cite{Per}. The fiber over the base point $\hat{\rho}(I)=\pi(I)$ is precisely the set $\Oo_I(\Hh'_\RM,J)$ from \eqref{eq-OQdef}. Hence one can use the long exact sequence of homotopy groups, which due to the triviality of the homotopy groups of $\Oo(\Hh'_\RM,J)$  implies that the set $\Omega_{\hat{\rho}(I)}\Cc(\Hh'_\RM)$ of based loops in the base space is homotopy equivalent to the fiber over the base point which here is $\Oo_I(\Hh'_\RM,J)$. Because the loop functor respects homotopy, we conclude from the above that the based loop space $\Omega_I\hat{\Ff}^0$ is homotopy equivalent to $\Oo_I(\Hh'_\RM,J)$. The last claim follows from $\pi_1(\hat{\Ff}^0)\cong \ZM_2$.
\hfill $\Box$

\vspace{.2cm}

It is possible to use the index map $j_I:\Oo_I(\Hh'_\RM,J)\to\ZM_2$ defined by
$$
j_I(O)
\;=\;
\PI (I,OIO^*)
$$
to distinguish the two components of $\Oo_I(\Hh'_\RM,J)$. Furthermore, applying Theorem~\ref{theo-jproperties} and Proposition~\ref{theo-Pf2j} to $I_0=I$ and $I_1=OIO^*$ leads to the following.

\begin{coro} 
\label{coro-Pf2j}
For any chiral complex structure $I$,  $j_I$ is a homotopy invariant homomorphism labelling the two components of $\Oo_I(\Hh'_\RM,J)$. One has
$$
j_I(O)
\;=\;
\SF_2\big([0,1]\ni t\mapsto (1-t)\,I\,+\,t\,OIO^*\big)
\;.
$$
\end{coro}

The (Noether) index of a Toeplitz operator associated to a given index pairing can always be expressed as a spectral flow  \cite{Ph,Ph1,DS2}. The following result is the parity version of this result, similar to \cite{CPS} which contains a corresponding result for the $\ZM_2$-valued spectral flow.

\begin{theo} 
\label{theo-index}
Let $I$ be a real chiral complex structure on $\Hh'_\RM$ and $O\in\Oo_I(\Hh'_\RM,J)$. If $P$ is the spectral projection onto the positive imaginary spectrum of $I$ where $I$ was extended to a skew-adjoint operator on $\Hh'_\RM \otimes \CM$, then
$$
\SF_2\big([0,1]\ni t\mapsto (1-t)I+tOIO^*\big)
\;=\;
\dim_\CM\big(\Ker_\CM(POP+\one-P)\big)\;\mbox{\rm mod} \,2
\;.
$$
\end{theo}

\noindent {\bf Proof.} First of all, the $\ZM_2$-index on the right hand side is of the type $(j,d)=(1,8)$ in Theorem~1 of \cite{GS}.  Indeed, $P$ satisfies $\overline{P}=\one-P$ and $JPJ=\one-P$ (namely, $P$ is even real and even Lagrangian in the terminology of \cite{GS}) as well as  $J\overline{O}J=O$. In particular, the index pairing on the right hand side is a homotopy invariant under variations of $O$ and $P$ respecting all the properties mentioned above. Now given $I$, the set $\Oo_I(\Hh'_\RM,J)$ has two components by Theorem~\ref{theo-IsoQqQ}. The proof of Theorem~\ref{theo-index} is thus remarkably simple. Both sides of the equality are homotopy invariants and lie in $\ZM_2$. Hence it is sufficient to verify equality on both components. For $O=\one$, both sides vanish. For the other component, the equality is verified for a non-trivial example in the next section.
\hfill $\Box$

\section{A non-trivial example}
\label{sec-example}

Let $\proj $ be a one-dimensional projection on an infinite-dimensional Hilbert space $\Hh_\RM$. We consider
$$
I\;=\;\begin{pmatrix} 0 & \one \\ -\one & 0 \end{pmatrix}
\;,
\qquad
O\;=\;\begin{pmatrix} \one-2\proj  & 0 \\ 0 & \one \end{pmatrix}
\;.
$$
Then all conditions in Theorem~\ref{theo-index} are satisfied. One has
$$
P\;=\;
\frac{1}{2}  \begin{pmatrix} \one & -\imath\one \\ \imath\one & \one \end{pmatrix}
\;,
$$
and
$$
POP+(\one-P)
\;=\;
\one\,-\,\frac{1}{2}  \begin{pmatrix} \proj  & -\imath\proj  \\ \imath\proj  & \proj  \end{pmatrix}
\;.
$$
In particular, $\dim(\Ker(POP))=\dim(\proj )=1$. Hence the index on the right hand side of Theorem~\ref{theo-index} is equal to $1$. On the other hand, the straight-line path is
$$
I_t\;=\;
(1-t)\,I\,+\,t\,O^*IO
\;=\;
\begin{pmatrix} 0 & \one-2\,t\,\proj  \\ -\one+2\,t\,\proj  & 0 \end{pmatrix}
\;.
$$
Hence this contains exactly one copy of the example \eqref{eq-examp}, and so $\SF_2([0,1]\ni t\mapsto I_t)=-1$  (notably, the non-trivial value).

\vspace{.2cm}

The above path can be completed to a loop with $[1,2]\ni t\mapsto I_t=O_t^*I O_t$ where $[1,2]\ni t\mapsto O_t$ is a Kuipers path connecting $O$ to $\one$. As this second path is in the invertibles it has trivial $\ZM_2$-valued spectral flow. Therefore $[0,2]\ni t\mapsto I_t$ is a loop with non-trivial $\ZM_2$-valued spectral flow. This provides the example needed in the proof of Theorem~\ref{theo-isomorphism}.

\vspace{.2cm}

Let us also calculate the parity of $[0,1]\ni t\mapsto I_t$ as in \cite{FP}. One needs to look at the off-diagonal entry $B_t$ as in \eqref{eq-link}, and determine an invertible operator $M_t$ such that $M_tB_t-\one=K_t$ is compact. Clearly, $M_t=\one$ will do, and then $K_0=0$ and $K_1=-2\proj $ so that $\degLS(T_0)=1$ and $\degLS(T_1)=-1$. Thus one finds again that the parity of the path is $-1$.

\section{Application to a topological insulator}
\label{sec-TopIns}

In the following the reformulation with chiral self-adjoints from Chapter~\ref{sec-Extensions} is used. Let $\Hh_\CM=\ell^2(\ZM)\otimes \CM^N$ and consider the following operator on $\Hh'_\CM=\Hh_\CM\oplus\Hh_\CM$:
$$
H_t
\;=\;
\begin{pmatrix}
0 & (S_t)^k\otimes\one_N \\
(S_t^*)^k\otimes\one_N & 0
\end{pmatrix}
\;,
$$
where $k\in\ZM$ and $S_t$ is the bilateral shift perturbed on one link from $1$ to $\cos(\pi t)$, namely in Dirac notation
$$
S_t\;=\;\sum_{n\not=0}|n\rangle\langle n+1|\;+\;\cos(\pi t)\,|0\rangle \langle 1|
\;.
$$
The Hamiltonian has the chiral symmetry \eqref{eq-ChiralSym} and is real as well as self-adjoint for all $t$: 
$$
H_t\;=\;H_t^*\;=\;\overline{H_t}\;=\;-JH_tJ
\;.
$$
Thus it is possible to consider the parity of the path $[0,1]\ni t\mapsto H_t$. One finds
$$
\PF([0,1]\ni t\mapsto H_t)
\;=\;
(-1)^{kN}
\;.
$$
This property is now stable under any kind of perturbations not closing the spectral gap of $H_0$, such as a chiral disordered potential $V_\omega=-JV_\omega J$ of moderate strength. Here $\omega$ is a point in a compact W$^*$-dynamical system $(\Omega,T,\ZM,\PM)$ given by the shift action $T$ of $\ZM$ and an invariant and ergodic probability measure $\PM$. Let us comment that the non-triviality of the path $[0,1]\ni t\mapsto H_t$ has nothing to do with the strong invariants appearing in the periodic table of topological insulators. The Hamiltonian has an even time-reversal symmetry and a chiral symmetry. Hence  it lies in the so-called BDI class. As such, in dimension $d=1$ there are infinitely many distinct phases labelled by the strong invariant, which in the above example is the number $kN$ specifying the winding of the off-diagonal entry of $H_0$. For each $k$, the Hamiltonian is then in the corresponding component of Fredholm operators and stays within it along the path $[0,1]\ni t\mapsto H_t$, because it resulted from a merely local perturbation of $H_0$. It is now a fact that such paths can be topologically non-trivial because the fundamental group of $\Ff^0$ is $\ZM_2$. The parity detects this topology.

\vspace{.2cm}

Let us now come to the physical implications of the non-trivial parity. One can directly conclude that $H_t$ has to have an eigenvalue crossing through $0$ at some $t\in[0,1]$. However, more can be said, namely one such eigenvalue crossing has to take place at half flux. 

\begin{theo} 
\label{theo-ZeroMode} If $kN$ is odd, then $H_{\frac{1}{2}}$ has an odd number of evenly degenerate zero modes, namely the multiplicity of $0$ as eigenvalue is $2$ modulo $4$.
\end{theo}

\noindent {\bf Proof.} Let us introduce the gauge transformation
$$
G\;=\;\sum_{n>0}|n\rangle\langle n|\;-\;\sum_{n\leq 0}|n\rangle\langle n|
\;.
$$
Then $GS_tG=S_{1-t}$ so that $GH_tG=H_{1-t}$. Consequently the zero eigenvalue crossings for $t$ lead to zero eigenvalue crossings for $1-t$. As the parity is invariant under a change of orientation and also unitary conjugations, these eigenvalue crossing cancel and do not lead to a net parity, except at $t=\frac{1}{2}$. This implies that at $t=\frac{1}{2}$ one has to have an odd number of eigenvalue crossings. Consequently, the multiplicity of the zero eigenvalue is $2$ modulo $4$. 
\hfill $\Box$

\vspace{.2cm}

Note that at half-flux, the shift does not connect left and right half-space so that $H_{\frac{1}{2}}$ is a diret sum of a left and a right half-space Hamiltonian. Each has to have a zero mode, leading to the two-fold symmetry. Let us further add a few comments on how to interpret Theorem~\ref{theo-ZeroMode} against the background of the periodic table of topological insulators. As already stated, all the above Hamiltonians are from the BDI class of chiral Hamiltonians with an even time-reversal symmetry (integer spin). Also the Hamiltonian $H_{\frac{1}{2}}$ is within this class. If one extracts only the low lying spectrum (eigenvalues in the vicinity of $0$), this reduced Hamiltonian is a finite dimensional matrix and hence represents a system of dimension $d=0$ (corresponding to the local defect induced by a half-flux). The set of $0$-dimensional BDI Hamiltonians has two components which are distinguished by the parity of the zero modes (of each half-sided Hamiltonian). Theorem~\ref{theo-ZeroMode} states that $H_{\frac{1}{2}}$ is in the non-trivial component of the $0$-dimensional BDI Hamiltonians always having a zero mode.

\section{Application to bifurcation theory}
\label{sec-Bifur}

The aim of this final section is to apply the parity in the bifurcation theory of solutions to nonlinear operator equations depending on a real parameter. It was precisely for this purpose that the parity was originally introduced and put to work \cite{FP0,FP1,FP}. The treatment given in this note suggests to construct examples with a skew-adjoint linearization which has a chiral symmetry built in. This is essentially what is done below.

\vspace{.2cm}

Let us begin by exposing the theoretical framework of bifurcation theory and the main result used later on. Given two real Banach spaces $X$ and $Y$ and an interval $I\subset\mathbb{R}$, one considers continuous maps $F:I\times X\rightarrow Y$ for which we assume throughout that $F(t,0)=0$ for all $t\in I$. In this context, one then calls the set $I\times\{0\}$ the trivial branch of solutions of $F(t, u)=0$. A bifurcation point for the family of equations $F(t,u)=0$, $t\in I$, is a parameter value $t^\ast$ where a new branch of solutions appears. 

\begin{defini}
A parameter value $t^\ast\in I$ is a bifurcation point for the family of equations $F(t,u)=0$ if in every neighborhood of $(t^\ast,0)\in I\times X$ there is some $(t,u)$ such that $u\neq 0$ and $F(t,u)=0$.
\end{defini} 

Let us now assume that the map $F$ is continuously differentiable in $u$. The implicit function theorem then implies that if the linear map $D_uF(t^\ast,0):X\rightarrow Y$ is invertible, there is a neighborhood of $(t^\ast,0)$ in $I\times X$ for which there is a unique solution of the equation $F(t,u)=0$. As $F(t,0)=0$ by assumption, we see that $t^\ast$ cannot be a bifurcation point in this case. Consequently, $D_uF(t^\ast,0)$ must be singular if $t^\ast$ is a bifurcation point. Let us stress, however, that not every $t^\ast$ for which $D_uF(t^\ast,0)$ is singular, is necessarily a bifurcation point. The aim of bifurcation theory is to find sufficient conditions under which a singular point $t^\ast$ is a bifurcation point. While such problems have been considered for centuries, topological criteria for existence of bifurcations in an infinite dimensional set-up were only made by Krasnoselskii in the sixties \cite{Kra}. 

\vspace{.2cm}

One extension of his ideas is the work of Fitzpatrick and Pejsachowicz \cite{FP1} which uses the parity and is described next. For a continuously differentiable $F$ let us consider the bounded linear operator $B_t=D_uF(t,0)$  and suppose that their index vanishes. As $ I\ni t\mapsto B_t$ is a continuous path by assumption, its parity is defined. 

\begin{theo}[\cite{FP1}]
\label{theo-Jacobo}
Suppose that $ I\ni t\mapsto B_t$ is an admissible path. If $\PF( I\ni t\mapsto B_t)=-1$, then there is a bifurcation point $t^\ast\in I$ for the family of equations $F(t,u)=0$. 
\end{theo}  

In the following, we provide an example of a parameter dependent system of partial differential equations for which the parity  can be calculated explicitly. On $\Omega=(0,\pi)\times(0,\pi)$ let us consider the family of elliptic systems parametrized by $t\in\RM$ 
\begin{equation}\label{equnonlin}
\left\{
\begin{aligned}
-\Delta u&\;=\; t v+f(t,x,u,v)\;,&& \,\text{in}\,\,\Omega\;,\\
-\Delta v&\;=\; t u+g(t,x,u,v)\;,&& \,\text{in}\,\,\Omega\;,\\
u&\;=\;v\;=\;0\;,&&\,\text{on}\,\,\partial\Omega\;,
\end{aligned}
\right.
\end{equation}  
where $u,v:\Omega\rightarrow\mathbb{R}$ and $f,g:\mathbb{R}\times\overline{\Omega}\times\mathbb{R}^2\rightarrow\mathbb{R}$ are continuously differentiable. We assume that $f(t,x,0,0)=g(t,x,0,0)=0$ for all $(t,x)\in\mathbb{R}\times\Omega$ so that $(u,v)=(0,0)$ is a solution of \eqref{equnonlin} for all $t\in\mathbb{R}$. Moreover, all partial derivatives of $f$ and $g$ with respect to $u$ and $v$ are supposed to be bounded and satisfy
\begin{align}\label{vanishingder}
D_{(u,v)} f(t,x,0,0)
\;=\;
D_{(u,v)}  g(t,x,0,0)\;=\;0
\;,
\qquad (t,x)\in\mathbb{R}\times\Omega
\;.
\end{align}
As the Laplacian  as operator on $L^2(\Omega)$ with domain $H^2(\Omega)\cap H^1_0(\Omega)$ is invertible with compact resolvent, one can transform the first two equations of \eqref{equnonlin} to the system
$$
F(t,u,v)
\;=\;
\begin{pmatrix}u\\v\end{pmatrix}+t\,\begin{pmatrix}Kv\\Ku\end{pmatrix}+\begin{pmatrix}K f(t,x,u,v)\\K g(t,x,u,v)\end{pmatrix}
\;=\;
0
\;,
$$
where $K=\Delta^{-1}:L^2(\Omega)\rightarrow L^2(\Omega)$ is compact. The assumptions on $f$, $g$ and the compactness of $K$ imply that $F:\mathbb{R}\times L^2(\Omega)\times L^2(\Omega)\rightarrow L^2(\Omega,\RM^2)$ is differentiable. Moreover, the derivative at $(0,0)$ of the nonlinear part vanishes by \eqref{vanishingder} and therefore
$$
B_t(u,v)
\;=\;
D_{(u,v)}F(t,0,0)(u,v)
\;=\;
\begin{pmatrix}u\\v\end{pmatrix}+t\,\begin{pmatrix}Kv\\Ku\end{pmatrix}
\;.
$$
By applying $\Delta$ to each component of the equation $B_t(u,v)=0$, one checks that 
$$
(u(x),v(x))
\;=\;
(\sin(x_1)\sin(x_2),\sin(x_1)\sin(x_2))
\;,
\qquad
x\,=\,(x_1,x_2)\,\in\,\Omega
\;,
$$
is in the kernel of $B_2$. To find out if $t^\ast=2$ is a bifurcation point of $F(t,u,v)=0$, let us now compute $\PF([2-\delta,2+\delta]\ni t\mapsto B_t)$. First of all, one needs to consider the eigenvalue problem 
$$
\begin{pmatrix}
0&B_t\\
-B^\ast_t&0
\end{pmatrix}
\begin{pmatrix}
z\\w
\end{pmatrix}
\;=\;
\mu\,
\begin{pmatrix}
z\\w
\end{pmatrix}
\;.
$$
By setting $z=(u_1,v_1)$ and $w=(u_2,v_2)$ and applying the Laplace operator in each component, this amounts to solve the system of equations 
\begin{equation}\label{equnlin}
\left\{
\begin{aligned}
\Delta u_2+ t v_2&\;=\;\mu\Delta u_1\;,&& \,\text{in}\,\,\Omega\;,\\
\Delta v_2+ t u_2&\;=\;\mu\Delta v_1\;,&& \,\text{in}\,\,\Omega\;,\\
-\Delta u_1- t v_1&\;=\;\mu\Delta u_2\;,&& \,\text{in}\,\,\Omega\;,\\
-\Delta v_1- t u_1&\;=\;\mu\Delta v_2\;,&& \,\text{in}\,\,\Omega\;,\\
u_1\;=\;u_2\;=\;v_1&\;=\;v_2\;=\;0\;,&&\text{on}\,\,\partial\Omega\;.
\end{aligned}
\right.
\end{equation}
Setting for integer $k_j,m_j,l_j,n_j$ where $j=1,2$, 
\begin{align*}
u_1(x)&\;=\;\sin(k_1x_1)\sin(k_2x_2),\qquad u_2(x)\;=\;\sin(m_1x_1)\sin(m_2x_2)\;,\\
v_1(x)&\;=\;\sin(l_1x_1)\sin(l_2x_2),\qquad \;v_2(x)\;=\;\sin(n_1x_1)\sin(n_2x_2)\;,
\end{align*}
the equations \eqref{equnlin} are equivalent to    
\begin{equation*}
\left\{
\begin{aligned}
-(m^2_1+m^2_2) u_2+ t v_2&\;=\;-\mu(k^2_1+k^2_2) u_1\;,&& \,\text{in}\,\,\Omega\;,\\
-(n^2_1+n^2_2) v_2+ t u_2&\;=\;-\mu(l^2_1+l^2_2) v_1\;,&& \,\text{in}\,\,\Omega\;,\\
(k^2_1+k^2_2) u_1- t v_1&\;=\;-\mu(m^2_1+m^2_2) u_2\;,&& \,\text{in}\,\,\Omega\;,\\
(l^2_1+l^2_2) v_1- t u_1&\;=\;-\mu(n^2_1+n^2_2) v_2\;,&& \,\text{in}\,\,\Omega\;.\\
\end{aligned}
\right.
\end{equation*}
It is readily seen that for $t$ close to $2$, one can only have an eigenvalue crossing zero in the subspace of $L^2(\Omega,\mathbb{C}^2)\oplus L^2(\Omega,\mathbb{C}^2)$ spanned by $(u_1,0,0,0)$, $(0,v_1,0,0)$, $(0,0,u_2,0)$ and $(0,0,0,v_2)$ when $k_1=k_2=m_1=m_2=l_1=l_2=n_1=n_2=1$. The four eigenvalues in this subspace are
$$
\lambda_1
\;=\;
-\tfrac{\imath}{2}(t-2)
\;,
\quad 
\lambda_2\;=\;\tfrac{\imath}{2}(t-2)\;,
\quad 
\lambda_3\;=\;-\tfrac{\imath}{2}(t+2)\;,
\quad 
\lambda_4\;=\;\tfrac{\imath}{2}(t+2)
\;.
$$
The eigenvalue crossing is simple and analytic, and of the type of the first example in \eqref{eq-examp}. Consequently, $\PF([2-\delta,2+\delta]\ni t\mapsto B_t)=-1$ for all small $\delta>0$ and thus  $t^\ast=2$ is indeed a bifurcation point for \eqref{equnonlin} by Theorem \ref{theo-Jacobo}.

\vspace{.2cm}

\noindent {\bf Acknowledgements:} This work was partially supported by the DFG. N.W. thanks the Friedrich-Alexander-Universtit\"at for a Visiting Professorship.

\vspace{-.2cm}


\end{document}